\documentclass[runningheads]{llncs}
\usepackage[T1]{fontenc}
\usepackage{amsfonts}
\usepackage{amsmath}
\usepackage{tikz}
\usepackage{caption}
\usepackage{subcaption}
\usepackage{float}
\usepackage{placeins}
\usepackage{svg}
\usepackage{layouts}
\usepackage{float}

\usetikzlibrary{positioning}
\usetikzlibrary{backgrounds}
\usetikzlibrary{arrows, arrows.meta}

\usepackage{graphicx}
\usepackage{hyperref}
\usepackage{cite}
\usepackage{color}

\begin{document}
\title{Cell segmentation of \textit{in situ} transcriptomics data using signed graph partitioning}
\titlerunning{Cell segmentation of IST data using signed graph partitioning}
\author{Axel Andersson* \and
Andrea Behanova* \and
{Carolina Wählby} \and
{Filip Malmberg}}

%

\authorrunning{A. Andersson et al.}
%
\institute{{Centre for Image Analysis, Department of Information Technology and SciLifeLab BioImage Informatics Facility, Uppsala University, Sweden}\\
$*$ These authors contributed equally\\
\email{\{axel.andersson, andrea.behanova, carolina.wahlby, filip.malmberg\}@it.uu.se}}
\maketitle              
\begin{abstract}
The locations of different mRNA molecules can be revealed by multiplexed in situ RNA detection. By assigning detected mRNA molecules to individual cells, it is possible to identify many different cell types in parallel. This in turn enables investigation of the spatial cellular architecture in tissue, which is crucial for furthering our understanding of biological processes and diseases. However, cell typing typically depends on the segmentation of cell nuclei, which is often done based on images of a DNA stain, such as DAPI. Limiting cell definition to a nuclear stain makes it fundamentally difficult to determine accurate cell borders, and thereby also difficult to assign mRNA molecules to the correct cell. As such, we have developed a computational tool that segments cells solely based on the local composition of mRNA molecules. First, a small neural network is trained to compute attractive and repulsive  edges between pairs of mRNA molecules. The signed graph is then partitioned by a mutex watershed into components corresponding to different cells. We evaluated our method on two publicly available datasets and compared it against the current state-of-the-art and older baselines. We conclude that combining neural networks with combinatorial optimization is a promising approach for cell segmentation of in situ transcriptomics data. The tool is open-source and publicly available for use at \url{https://github.com/wahlby-lab/IS3G}.

\end{abstract}

\section{Introduction}
Over the past years, a large number of techniques for spatially resolved multiplexed in situ transcriptomics (IST) have been developed~\cite{Ke2013, Moffitt2018, Codeluppi2018, Eng2019, directRNA, xenium}. These techniques enable the mapping of hundreds of different mRNA molecules directly within tissue samples, allowing the dissection and analysis of cell type heterogeneity while preserving spatial information. These techniques produce large gigapixel-sized images of tissue sections with millions of different spatial biomarkers for various mRNA molecules. A single experiment can pinpoint the location of hundreds of different types of mRNA molecules with sub-micrometer resolution. The many different types of targeted mRNA molecules, as well as a large number of detected molecules, make visual exploration and analysis challenging. To alleviate, and compute quantitative statistics, a range of tools, as described below, have been developed for grouping the mRNA molecules into groups, such as cells, cell types, or tissue-level compartments. 

\subsection{Tools for analyzing IST data}
The analysis of IST data typically starts by assigning mRNA molecules to localized cells.
By examining the composition of molecules within cells, it becomes possible to define and analyze different cell types and their spatial organization~\cite{scanpy, squidpy, seurat, FICT}. Traditionally, cells are located in a nuclear stained image complementary to the IST experiment using techniques such as Stardist~\cite{stardist}, Cellpose~\cite{cellpose}, or the distance-transform watershed approach. Subsequently, the mRNA molecules are assigned to the detected nuclei based on their proximity in space. However, assigning molecules to cells solely based on their spatial proximity to the nucleus may not be optimal due to irregular cell shapes and the asymmetric distribution of detected mRNAs around the nuclei.  Moreover, there are situations where clusters of mRNA molecules belonging to a cell are detected in the IST experiment, but the corresponding nucleus lies outside the imaged region of interest. Additionally, if the quality of the nuclear image is poor and the nuclei are not clearly visible, it further complicates the assignment process. In such cases, cell detection based on nuclear staining is not optimal.

To improve the assignment of molecules to already detected cells, Qian et al~\cite{Qian2019} created a probabilistic cell typing method, pciSeq, where prior known information regarding the molecular composition of different cell types is utilized when assigning molecules and typing cells. Similarly, Prabhakaran et al.~\cite{prabhakaran2022sparcle} introduced Sparcle, a method where mRNAs are assigned to cells through a relatively simple "assign" and "refine" algorithm. However, both Sparcle and pciSeq require that the location of cells is known beforehand. Petukhov et al.~\cite{Petukhov2020} introduced Baysor, an extensive probabilistic model that both detects the location of cells and assigns molecules to the cell. Alternatively, there are methods that overcome assigning molecules to cells by simply ignoring the cells, and instead assigning molecules to spatial bins~\cite{Park2021,Tiesmeyer2022,partel2020automated}, or using deep learning to learn more abstract features~\cite{Partel2020,hu2021spagcn,ccst}. Such methods allow the user to easily identify regions of similar molecular compositions (corresponding to semantic segmentation), but statistics on a per-cell level (requiring instance segmentation) are difficult to compute.

\subsection{Contribution}
Localizing cells and assigning the right molecules to cells is a crucial task in IST. In this context, we introduce In Situ Sequencing SEGmentation (IS3G), a novel tool that jointly identifies the location of cells and assigns mRNA molecules to them, without the need for prior knowledge of cell location or cell type molecular compositions. IS3G operates solely on local mRNA composition and identifies cells by partitioning a signed graph, making it possible to segment cells without relying on nuclear staining.  Our results demonstrate that signed graph partitioning can be used to efficiently segment IST data without seeds for cell location or cell types.

\section{Methodology}
In brief, IS3G utilizes a small neural network to classify whether two mRNA molecules originate from the same cell or not.  The posterior probabilities of the classifier are used to determine the strength of attractive (positive) and repulsive (negative) edges in a signed graph. Using a mutex watershed~\cite{mutex}, this graph is then partitioned into components that correspond to individual cells, enabling the accurate assignment of molecules to their respective cells.

\subsection{Compositional features}
The neural network used to predict the attractive and repulsive edge strengths is trained on local compositions of mRNA molecules. In this section, we explain how to extract such  features. We start by setting the notation. In an IST experiment, the $i$'th detected mRNA molecule can be described with two attributes: a position $p_i \in \mathbb{R}^D$ and a categorical label $l_i \in \mathbb{L}$, where $D$ is the number of spatial dimensions (usually two or three) and $\mathbb{L}$ is the set of targeted mRNA molecules. The categorical label of a molecule can be represented as a one-hot-encoded vector, i.e., $e_i \in \{0,1\}^{|\mathbb{L}|}$. The local composition of mRNA molecules in a neighborhood around the $i$'th mRNA can thus be computed by simply counting the labels among the $k$ nearest neighbors,
\begin{equation}
x_i = \sum_{j \in \mathcal{N}(p_i)} e_j,
    \label{eq:1}
\end{equation}
where $\mathcal{N}(p_i)$ refers to the $k$ nearest neighbors to the $i$'th molecule. The parameter $k$ depends on the molecular density and must be tuned so that the compositional features describe molecular patterns on a cellular scale. We found that setting the value of $k$ to roughly one third of the expected molecular count per cell works well. The compositional features $x_i$ will serve as the input to our model used for computing the strength of our attractive and repulsive edges.

\begin{figure}[hbt!]
    \centering
    \includegraphics[width=0.96\textwidth]{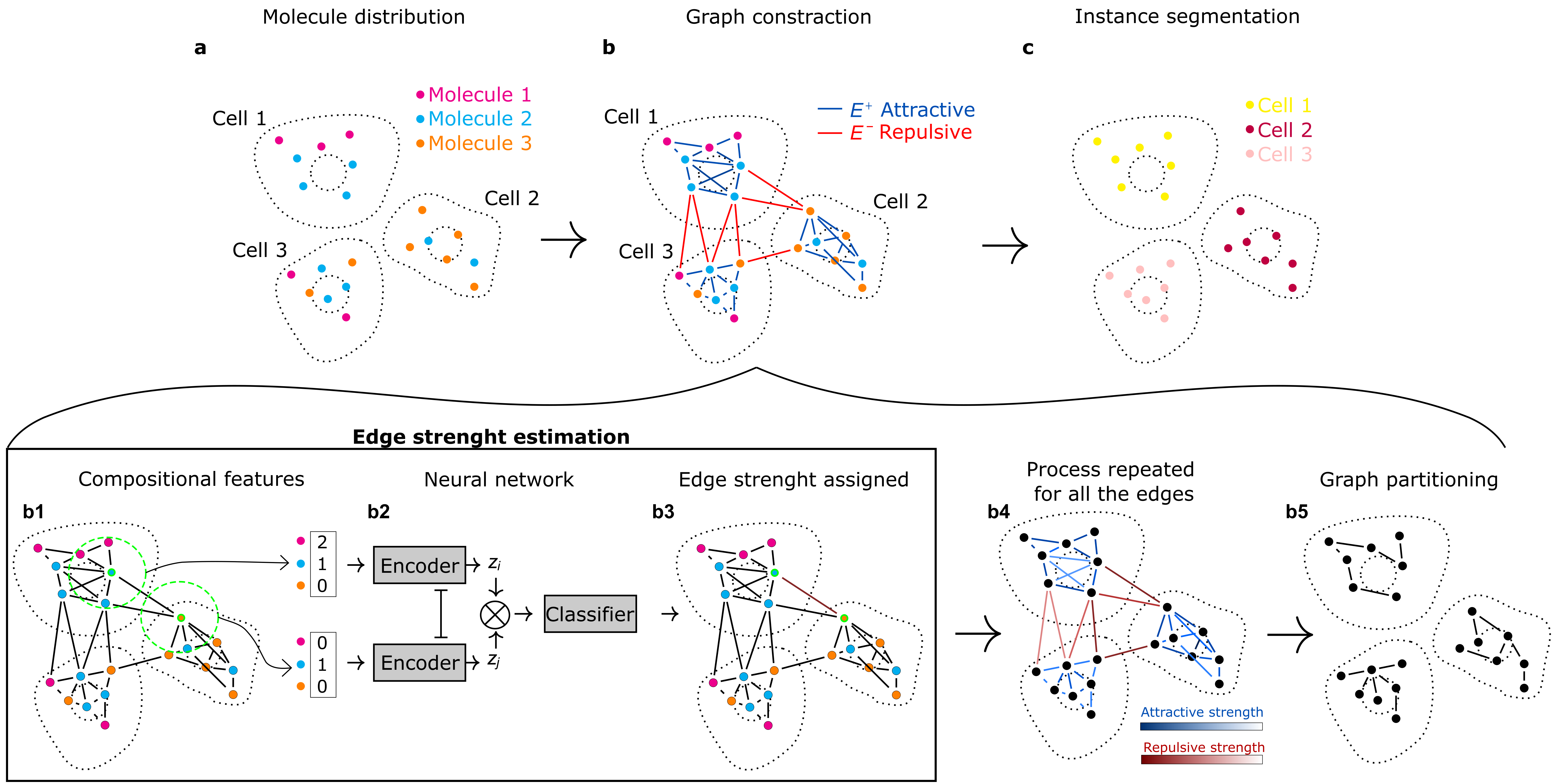}
    \caption{Explanatory overview of the method. The top row: (\textbf{\textsf{a}}) Visualization of the distribution of molecules within three cells, where each color represents a different type of mRNA molecule. (\textbf{\textsf{b}}) Attractive edges connect molecules over short distances, whereas repulsive edges connect molecules over greater distances. The weight of the attractive edges indicates the \textit{desire} that the two molecules belong to the same cell, whereas the weight of the repulsive edges indicates a desire that the molecules belong to two different cells. Panel (\textbf{\textsf{b1-b3}}) describes how the edge weights are computed. First, compositional vectors are computed for a pair of molecules (\textbf{\textsf{b1}}). The pair of compositional vectors, ($x_i$, $x_j$), are respectively fed through an encoder, transforming the vectors into latent representations ($z_i$, $z_j$) (\textbf{\textsf{b2}}). The Euclidean distance between the latent vectors is fed through a classifier, attempting to classify whether the pair of latent vectors are from the same cell or not. The posterior probabilities of the classifier are used to set the strength of the attractive and repulsive edges (\textbf{\textsf{b3}-\textsf{b4}}). The signed graph is partitioned using a mutex watershed. The connected components defined by the active attractive edges are labeled individual cell instances (\textsf{\textbf{b5,c}}.)}
    \label{fig:Method_overview}
\end{figure}

To predict the edge strengths between molecules, we train a small Siamese neural network to classify whether two molecules belong to the same cell or not. The network consists of an encoder  and a classifier. First, the compositional features are computed for each of the pairs according to Eq.~\ref{eq:1}. The pair of features, $(x_i, x_j)$, are respectively encoded by the encoder into latent vectors $(z_i, z_j)$ using a simple neural network with two fully-connected layers with ELU activations. Next, the Euclidean distance between the two latent vectors is computed, $\Delta_{i,j} = \|z_i - z_j \|_2$. Based on the distance between the vectors, the network attempts to classify whether the sampled molecule pair belongs to the same cell or not. The classifier consisted of a single fully connected layer ending with a Sigmoid function. The posterior probability, $y_{ij}$, is then used as the attractive and repulsive ($y_{ij}-1)$ edge strengths. However, compositional features ($x_i$) computed in regions with low concentrations of mRNA molecules are inherently more sensitive to small molecular variations than compositional features computed in high-concentration regions. We, therefore, scale edge weights by the density factor $\rho_{ij} = \mathrm{min}(\|x_i\|_1,\|x_j\|_1)$ to give edges between pairs of mRNA molecules in low concentration regions lower weight than pairs in high concentration regions.

The weights of the encoder and classifier are jointly optimized by minimizing the binary cross entropy using the Adam optimizer~\cite{adam}. Training data is generated stochastically using the heuristic that two molecules separated by a distance less than $R_{\mathrm{cell}}$ are labeled as belonging to the same cell, whereas molecules separated by a distance larger than $2R_{\mathrm{cell}}$ are labeled to belong to two different cells. We trained the network for a maximum of 300 epochs, where one epoch was reached when we sampled the same number of molecule pairs as the total number of molecules in the dataset. However, 300 epochs were never reached since we employed an early stopping procedure. If the loss was non-decreasing for more than 15 epochs, the training would terminate. The model with the lowest loss was saved and used for computing edge strengths. A batch size of 2048 was used for all experiments.

\subsection{Graph construction and partitioning}
In the previous section, we explained how a neural network can be used to compute attractive and repulsive edge strengths. In this section, we will define our set of attractive and repulsive edges. We define our set of attractive edges, $E^+$, by assuming that molecules separated by a short distance are likely to belong to the same cell. Specifically, $E^+$ is created by connecting each molecule to all of its five nearest neighbors.

Next, we define our set of repulsive edges between molecules separated by a distance larger than $2R_{\mathrm{cell}}$ and less than $6R_{\mathrm{cell}}$. To save memory, each molecule is only connected with a repulsive edge to $15$ randomly selected neighbors within the interval. The repulsive edge weights are set to $-\infty$ for edges between molecules separated by a distance larger than $4R_{\mathrm{cell}}$. 

Finally, we use the procedure from the previous section to compute the strength of the edges, leaving us with a graph with signed edge weights. The graph is finally partitioned using a mutex watershed into components corresponding to different cells. 

The methodology is shown as an illustrative example in Fig.~\ref{fig:Method_overview}.

\subsection{Pre and Post processing}
To speed up the segmentation we first remove markers in low-density regions, as these markers are likely extracellular. We identify these markers automatically by computing the distance to the 15'th nearest neighbor. The distance is then clustered using a Gaussian mixture model. Markers belonging to the component with the larger mean are deemed extracellular and removed. After the segmentation, we discard cells with fewer than $n_{\mathrm{min}}$ number of molecules.  

\subsection{Visualization}
The segmented cells can be visualized in two ways: by assigning a color to each marker based on the cell it belongs to, or by outlining the segmented cell with a contour. To generate the contours, we employ an algorithm similar to the one described in~\cite{Petukhov2020}. For each cell, we calculate the marker frequency within small spatial bins placed on a regular grid. This process is repeated for all detected cells, resulting in a multi-channel count image. Each pixel in the image represents the marker count in a bin for a specific cell. The count image is spatially smoothed using a Gaussian filter and then transformed into a 2D labeled mask by identifying the index of the maximum value across the channels.  Each label in the mask thus corresponds to detected cells. Next, we find the contours of each cell in the labeled mask. The contour pixels for each cell are finally filtered by taking the longest path of the minimum spanning tree connecting the points. 

\section{Experiments}
We perform experiments on two publicly available datasets: An osmFISH~\cite{Codeluppi2018} and an In Situ Sequencing (ISS)~\cite{Qian2019} dataset and compare with authors' original segmentation as well as Baysor~\cite{Petukhov2020} --- the current state-of-the-art.

\subsection{osmFISH}
\begin{figure}
    \centering
    \includegraphics[trim={0.05cm 4.2cm 0cm 0cm},clip]{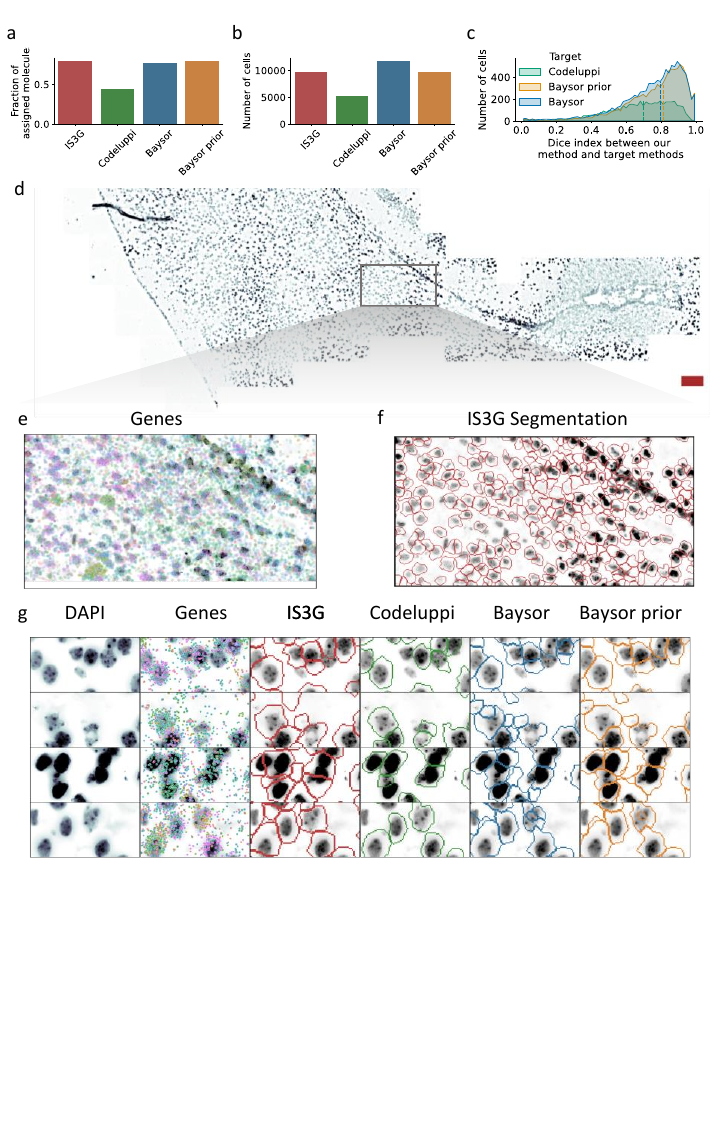}
    \caption{Results of various segmentation techniques applied on osmFISH dataset. 
    The total number of detected cells and the fraction of molecules assigned cells are shown in \textbf{\textsf{a}} and \textbf{\textsf{b}} respectively. Cells detected by IS3G are paired with cells detected by other methods. The distribution of Dice indices of the paired cells is shown in \textbf{\textsf{c}}. The dashed line represents the median Dice index.  Panel \textbf{\textsf{d}} shows a zoomed out view of the DAPI image (100 $\mu$m scale bar), with zoom-ins showing the distribution of gene markers (\textbf{\textsf{e}}) and IS3G segmented cells (\textbf{\textsf{f}}). A series of segmentation examples are shown in \textbf{\textsf{d}}. Presented techniques are IS3G, Codeluppi, Baysor, and Baysor with DAPI. }    
    \label{fig:osmFISH_results}.
\end{figure}

We first studied the osmFISH dataset~\cite{Codeluppi2018}. This dataset consists of around two million molecules of 35 different types. Also included in this dataset is a segmentation produced by the original authors~\cite{Codeluppi2018}, here referred to as the Codeluppi method.  We ran IS3G using $R_{\mathrm{cell}} = 8~\mathrm{\mu m}$ and $k=35$. We also ran Baysor~\cite{Petukhov2020} on the dataset, using the parameters provided in their osmFISH example. Baysor can be seeded with prior information regarding the nuclei segmentation (as described in~\cite{Petukhov2020}), as such, we run Baysor both with and without such a prior. For each cell segmentation method, we filter out cells containing fewer than $n_{\text{min}} = 30$ molecules. First, we look at the number of cells detected by each of the methods as well as the fraction of assigned molecules. This is shown in~Fig.~\ref{fig:osmFISH_results}a and Fig.~\ref{fig:osmFISH_results}b respectively. As seen, IS3G finds roughly the same number of cells as Baysor and Baysor with prior, but significantly more than the original publication~\cite{Codeluppi2018}. 

Next, we wanted to investigate if we find cells in the same location as the other methods. To do this we first matched our detected cells with the other methods' detected cells based on Sørensen–Dice index (Dice index). If two cells identified by two methods contain exactly the same molecules, the Dice index is one, and zero if no molecules overlap between the two  cells. We match the cells between the two methods by maximizing the average Dice score across all detected cells using the Hungarian algorithm. Fig.~\ref{fig:osmFISH_results}c shows the distribution of Dice indices between matched cells detected using IS3G versus Baysor, Baysor with prior and Codeluppi et al.~\cite{Codeluppi2018}. The median Dice index between IS3G detected cells matched with Baysor detected cells was 0.8. Finally, Fig.~\ref{fig:osmFISH_results}d shows some examples of the segmentation done by the different methods. The full dataset with segmentation results from all the mentioned techniques can be found here:     \url{https://tissuumaps.scilifelab.se/osmFISH.html}
\begin{figure}
    \centering
    \includegraphics[trim={0 8.5cm 0 0cm},clip]{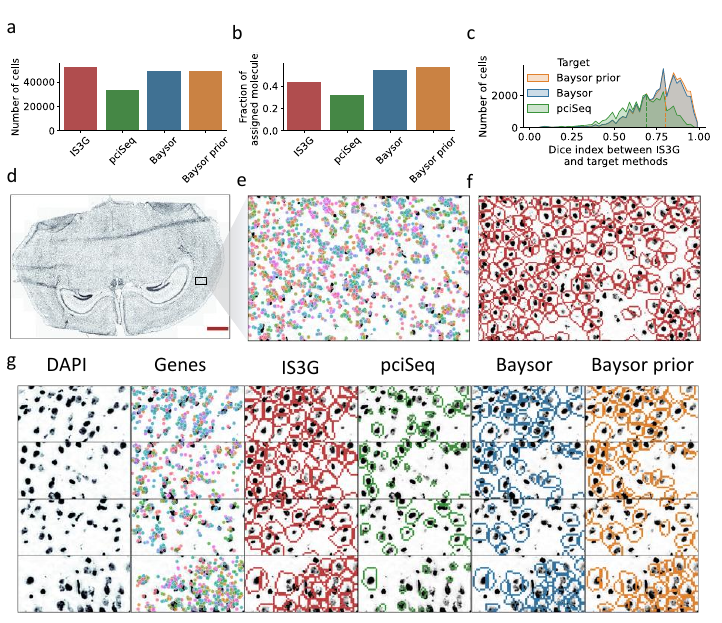}
    \caption{Results of various segmentation techniques applied on ISS dataset. 
    The total number of detected cells and the fraction of molecules assigned cells are shown in (\textbf{\textsf{a}}) and (\textbf{\textsf{b}}) respectively. Cells detected by IS3G are paired with cells detected by other methods. The distribution of Dice indices of the paired cells is shown in (\textbf{\textsf{c}}). The dashed line represents the median Dice index.  Panel (\textbf{\textsf{d}}) shows the DAPI image  of the whole data set (1 mm scale bar) with highlighted zoom-in sections showing gene markers (\textbf{\textsf{e}}) and IS3G segmented cells (\textbf{\textsf{f}}). A series of segmentation examples are shown in (\textbf{\textsf{g}}). Presented techniques are IS3G, pciSeq, Baysor, and Baysor with prior. }
    \label{fig:ISS_results}
\end{figure}
\subsection{In Situ Sequencing}
Secondly, we studied the dataset by Qian et al~\cite{Qian2019}. This dataset comprises around 1.4 million detected molecules of 84 different types. As for post-processing, we filter out cells containing fewer than $n_{\text{min}} = 8$ molecules. We used an $R_{\mathrm{cell}} = 10~\mu\text{m}$ and $k=8$. Fig.~\ref{fig:ISS_results}a shows the number of cells detected by IS3G, Baysor, Baysor with prior, and pciSeq. We note that IS3G finds roughly the same number of cells as Baysor and Baysor prior, and significantly more than pciSeq.

Fig.~\ref{fig:ISS_results}c shows the distribution of Dice indices when matching molecules assigned to our segmented cells with the other methods. The dashed lines indicate the median. Finally, Fig.~\ref{fig:ISS_results}d shows some examples of the segmentation done by the different methods. The full dataset with segmentation results from all the mentioned techniques can be found here: \url{https://tissuumaps.scilifelab.se/ISS.html}

\section{Discussion}
We have presented a simple technique for segmenting cells in IST data. IS3G differs from other approaches that need prior cell segmentation, seeds, or predetermined cell types. Instead, IS3G directly extracts features from the data by utilizing a simple neural network. We tested IS3G on two datasets, and it achieved performance comparable to the current state of the art, see Fig.~\ref{fig:osmFISH_results}c and Fig.~\ref{fig:osmFISH_results}c, showing that signed graph partitioning can be used to efficiently segment cells in IST data.

The deep learning model used to predict edge weights is very basic and likely not optimal. It was trained on "already aggregated data," which refers to compositional features obtained by computing the weighted frequency of molecules in circular neighborhoods, see Eq~\ref{eq:1}. A graph neural network may be more suitable for this application since it can also learn the weights used in the aggregation.

IS3G requires that the user provides a rough estimate of the cell radius, i.e., $R_{\mathrm{cell}}$. This parameter governs the bandwidth of the Gaussian kernel used when computing the compositional features. However, here we have assumed that the size of each cell is approximately the same. In practice, we have noticed that the size of the cells, or more precisely, the size of the mRNA point-cloud surrounding the cells, can vary between cells. Potentially, the segmentation could be improved by considering an adaptive bandwidth or using a graph-neural network that can extract features across multiple scales. 

 While not used explicitly herein, the mutex watershed algorithm provides a convenient way to specify mutually exclusive constraints between specific types of markers. This could be particularly beneficial in regions where it is difficult to differentiate between cells based solely on mRNA composition but a clear distinction can be made based on their nuclei. In such scenarios, IS3G may identify cells with multiple nuclei. However, if the user has supplementary markers indicating the position of nuclei, infinitely repulsive edges can be incorporated between these markers to explicitly ensures that each cell contains only one nucleus.

\bibliography{bibliography}
\bibliographystyle{splncs04}

\end{document}